\documentclass[fleqn,twoside,twocolumn,nofootinbib]{revtex4} 
\usepackage{ujp} 
\begin{document}
\title[Bistable organic materials]
{Bistable organic materials in optoelectrical switches:
Two-electrode devices vs organic field effect
transistors}%
\author{J. Sworakowski}
\affiliation{Institute of Physical and Theoretical Chemistry,
Wroc{\l}aw University of
Technology}
\address{Wyb. Wyspianskiego 27, 50-370 Wroc{\l}aw, Poland}
\email{sworakowski@pwr.wroc.pl}
\author{P. Lutsyk}%
\affiliation{Institute of Physical and Theoretical Chemistry,
Wroc{\l}aw University of
Technology}
\address{Wyb. Wyspianskiego 27, 50-370 Wroc{\l}aw, Poland}
\email{sworakowski@pwr.wroc.pl}
\affiliation{Institute of Physics, Nat. Acad. of
Sci. of Ukraine}%
\address{Prosp. Nauky 46, 03680 Kyiv, Ukraine}%
 \udk{???} \pacs{72.80.Le,  73.61.Ph,\\[-3pt]  81.05.Fb,
82.50.Hp, 85.30.Tv}

\razd{\secix}
\setcounter{page}{1021}%
\maketitle

\begin{abstract}
We present a short overview of research into properties of
organic materials and structures that could be used in
optoelectrical switches, i.e., switches in which changes in
electrical properties are triggered by light of appropriate
wavelengths. In particular, we describe the structures acting by
virtue of reversible photochemical reactions occurring in
photochromic molecular materials.
\end{abstract}

\section{Introduction}

One of the topics extensively studied in the domain of organic
electronics and photonics is the design and the fabrication of structures
that could serve as elements of memories and switches. According to
the textbook definition, a switch is a mechanical, electrical,
electronic, or optical device for opening or closing a circuit, or
for diverting the energy or charge from one part of a circuit to
another.

Figure 1 shows schematically the principle of action of a switch: the
structure, in its state ``1'' characterized by certain optical,
electrical, magnetic or mechanical parameters and stable for a
certain time interval, is converted by an activating stimulus to
state ``2'' (stable or metastable) characterized by another set of
parameters. The structure reverts to its initial state by a
deactivating stimulus. It is the matter of course that the hearts of
such structures should contain a multistable element. The activating
and deactivating stimuli can be optical, electrical, or magnetic
(e.g., the illumination with radiation triggering a chemical reaction,
electrical pulse switching the polarization of a ferroelectric
dielectric, pulse of a magnetic field switching the magnetization,
{\it etc.}

The recent paper by Heremans {\it et al.} [1] provides an extensive
review of electrically programmable memory devices. In this paper,
we present a short overview of research into properties of organic
materials and structures that could be used in optoelectrical
switches, i.e., switches in which changes in electrical properties
would be triggered by light of appropriate wavelengths. In
particular, we will concentrate on structures acting by virtue of
reversible photochemical reactions occurring in photochromic
molecular materials.

\begin{figure}[b]
\includegraphics[width=\column]{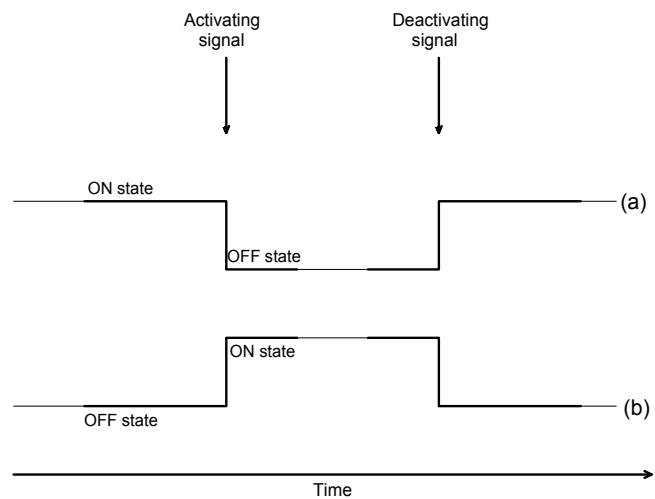}
\vskip-3mm\caption{ Schematic diagram illustrating the action of
switches: ({\it a}) ``normally ON''; ({\it b}) ``normally OFF'' }
\end{figure}

\begin{figure*}
\includegraphics[width=5.1cm]{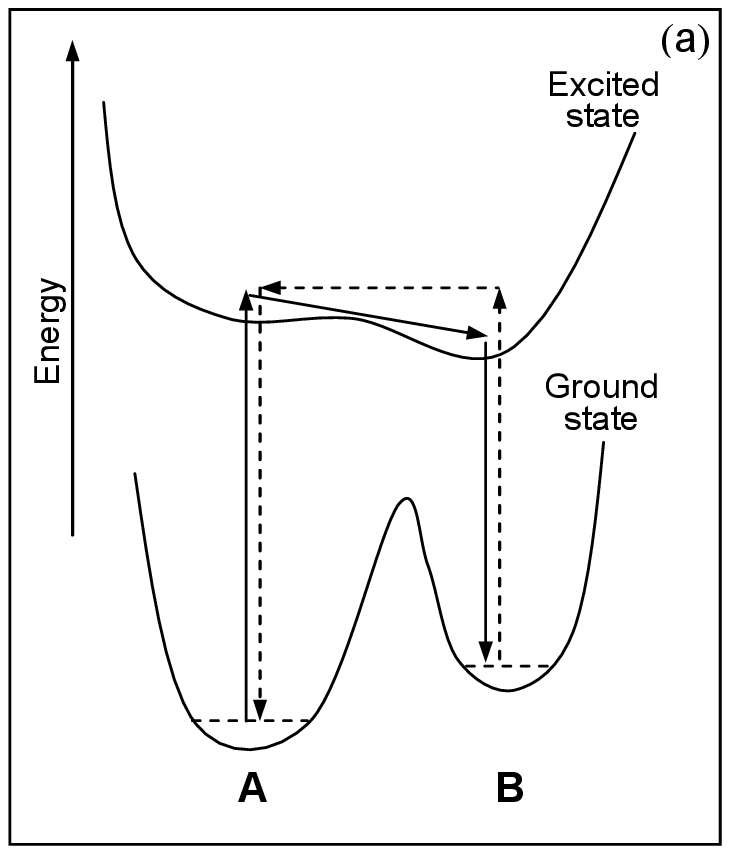}~~~~~~\includegraphics[width=5.0cm]{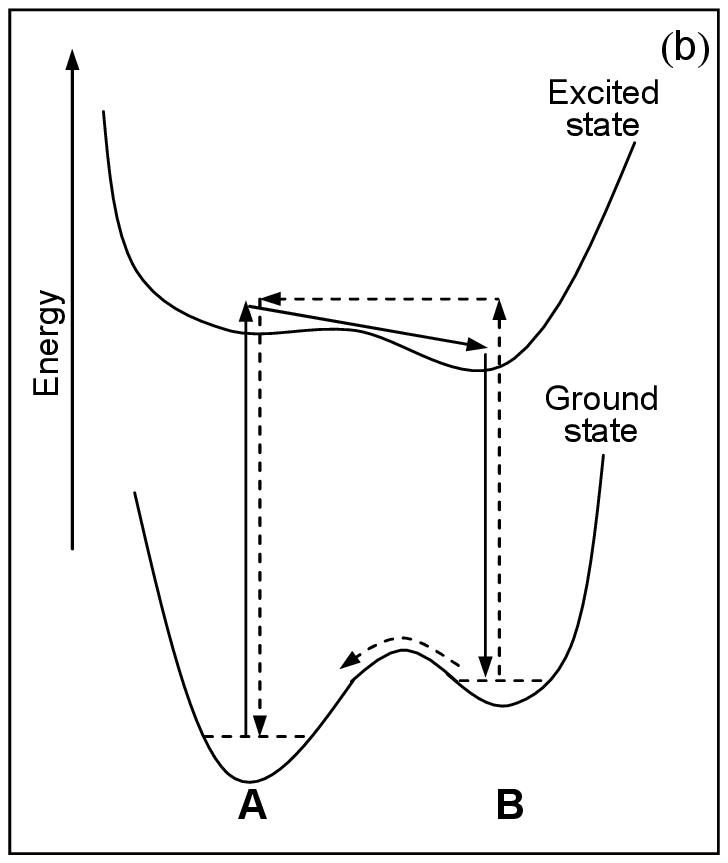}~~~\includegraphics[width=6cm]{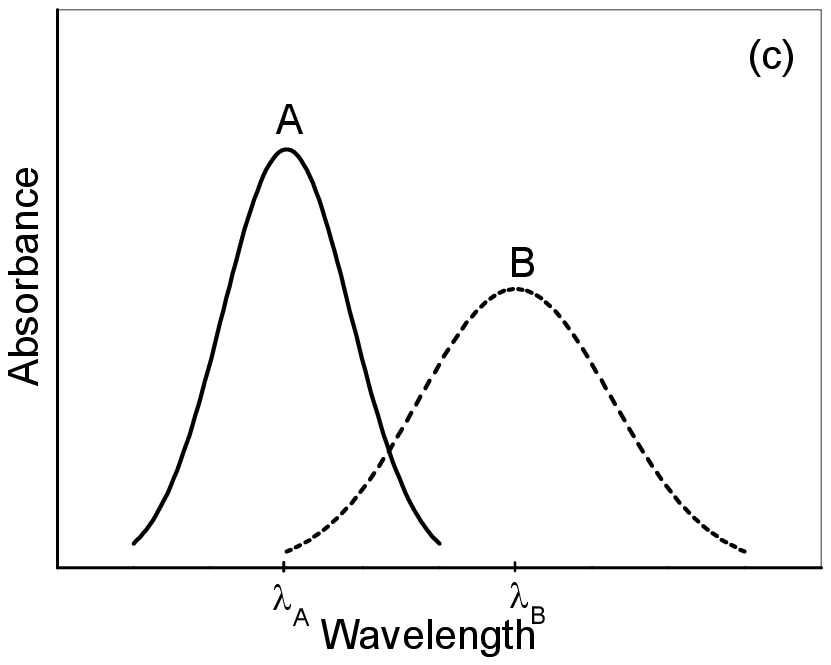}\\
\vskip-3mm\caption{ Schematic energy diagrams of the photochromic
switching process between reactant \textbf{A} and product
\textbf{B}. The \textbf{A} $\to $ \textbf{B} reaction is realized by
the optical stimulation via the excited electronic state. ({\it a}) High
energy barrier in the ground state between \textbf{B} and
\textbf{A}: the reverse reaction is possible only by the optical
excitation. ({\it b}) Low-energy barrier in the ground state between
\textbf{B} and \textbf{A}: the reverse reaction may occur both
photochemically (the same reaction path as in ({\it a}) and
thermally. ({\it c}) Absorption spectra associated with the
transitions between electronic levels in \textbf{A} and
\textbf{B} molecules }
\end{figure*}

The photochromic molecules are good examples of optically driven
bistable systems [2] that may allow one to control the current by
the illumination with light activating a reversible photochemical
reaction. The light of a given wavelength triggers the reaction by
changing electronic properties of a molecule (energies of electronic
levels, charge distribution on the molecule, {\it etc.}), while an
exposure to the radiation of a different wavelength results in a
reverse reaction, resulting in the return of the system to its
initial state (cf. Fig. 2). Suitably selected thermally stable
photochromic molecules are able to remain in their metastable form
for a sufficiently long time; hence, the current does not drop down
after turning off the illumination. The performance of the switch is
determined by the ratio of high-current (on) and low-current (off)
states (Fig. 1).

\section{Two-electrode Structures}

\subsection{Photochromic moieties molecularly dispersed in
organic semiconductor}

The idea of light-triggered electrical switching has been put
forward as early as a decade ago [3--6]. Early experiments carried
out on two-electrode structures (in sandwich or surface arrangement)
[7--11] contained photochromic systems and \textit{$\pi $}- or
\textit{$\sigma $}-conjugated semiconducting polymers.

Conceptually, the simplest way to achieve the effect is to graft a
suitable photochromic unit into the chain of a conjugated polymer.
This idea was developed by Kawai {\it et al}. [11] who incorporated
photochromic diarylethene directly into the main chain of a
semiconducting polymer (Fig. 3). The stable form of diarylethene
lacks the conjugation between the two moieties. Hence, the mean free path
of charge carriers should not exceed the average distance between
the incorporated photochromic groups. The conjugation path opens
under UV illumination, resulting in an increase of the mean free
path. Thus, electrically, with the stable form in the chain, the
system is in the ``OFF'' state, the UV illumination switches it to
the ``ON'' state, and the illumination with visible light,
triggering the reverse photochromic reaction, switches the molecule
back to the ``OFF'' state.

The system put forward by Sworakowski, Ne\v{s}purek, and coworkers
[3--8] also made use of a single-phase two-electrode system
comprising a conjugated polymer matrix and photochromic molecules,
either dissolved in the matrix or chemically attached to the polymer
chain as side groups (Fig. 4). The transport of charge carriers on
chains of a semiconducting polymer can be modulated by traps created
and annihilated by photochemically transformed photochromic
molecules. The mechanisms of trap formation slightly differ for
electron and hole transports [12]. As most organic semiconductors
are of the $p$-type, we will limit our discussion to the formation
of traps for holes.

One may envisage the formation of two types of traps by photochromic groups
attached to polymer chains or by photochromic molecules dissolved in organic
semiconducting (polymeric or low molecular weight) matrices: chemical traps
and dipolar traps. If the ionization energy of a photochromic species is
lower than that of the matrix, then chemical traps for holes are formed on
the photochromic dopant. The depth of such traps depends on the difference
of the ionization energies of the matrix ($I_h^g )$ and the dopant ($I_d^g )$.
Neglecting the second-order effects, the depth of a chemical trap may be written
as [12]
\begin{equation}
E_t^{\rm chem} \approx I_h^g -I_d^g,
\end{equation}
where the superscript $g$ indicates the ionization energy of an
isolated molecule.
The dipolar traps [13-16] are formed on molecules
in a vicinity of the dopant, if the dopant molecule is polar (i.e.,
possesses a non-zero permanent dipole moment), and the matrix is
either non-polar or weakly polar. In this case, the trap depth is
approximately equal to the interaction energy of the dipole moment
of the dopant and the carrier situated on an adjacent molecule of
the matrix. In the simplest case of a non-polar isotropic matrix,
the approximate expression reads
\begin{equation}
E_{id}=\frac{em\cos\Theta}{\varepsilon r^2},
\end{equation}
where $e$ is the unit charge, $m$ stands for the dipole moment of
the dopant, \textit{$\varepsilon $} is the relative electric
permittivity of the matrix, $r$ is the distance between the charge
and the dipolar dopant, and $\Theta $ is the angle between the
dipole moment and the radius vector of $r$. The trap depth is thus
proportional to the dipole moment of the dopant. More exact
calculations performed on a model molecular system [14, 15] showed
that traps as deep as ca. 0.6--0.7 eV can be created on a molecule
adjacent to a dipole (0.6 nm apart) whose electrical moment amounts
to 12 D. Experiments performed mainly of \textit{$\sigma
$}-conjugated polymer poly[(methylphenyl)silane] containing
photochromic spiropyran (Fig. 4) evidenced the opto-electrical
switching; the performance of two-electrode structures appeared,
however, moderate [7, 8]. As it seems, the unsatisfactory
performance of the switches results from important limitations that
must be overcome in practical realization. First of all, certain
relations between energy levels of the matrix and the photochromic
moieties have to be fulfilled [17]:

{\it i)} in order to prevent the simultaneous excitation of the matrix
and photochromic molecules, their excitation energies should be
separated ($\lambda _{\rm PC} >\lambda _{\rm matrix} )$;

{\it ii)} in order to prevent the energy transfer from the excited state
of photochromic molecules to the matrix, the energy of the
excited state of the matrix (measured with respect to a common
external reference) should be higher that that of the photochrom,
i.e., the following condition should be fulfilled:
\begin{equation}
\left ( I_{\rm PC}-\frac{hc}{\lambda_{\rm PC}} > I_{\rm
matrix}-\frac{hc}{\lambda_{\rm matrix}}\right).
\end{equation}
Moreover, a good performance of the device may only be achieved if
other important requirements are met:

{\it iii)} good chemical stability even in the short-wavelength part
of the spectrum (it would be better to realize the excitation with
light of low photon ener\-gy);

\begin{figure}
\includegraphics[width=8.0cm]{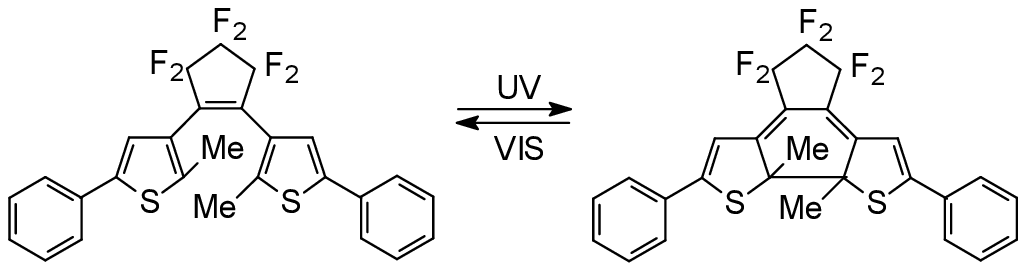}\\
{\large\it a}\\ \vskip2mm
\includegraphics[width=8.0cm]{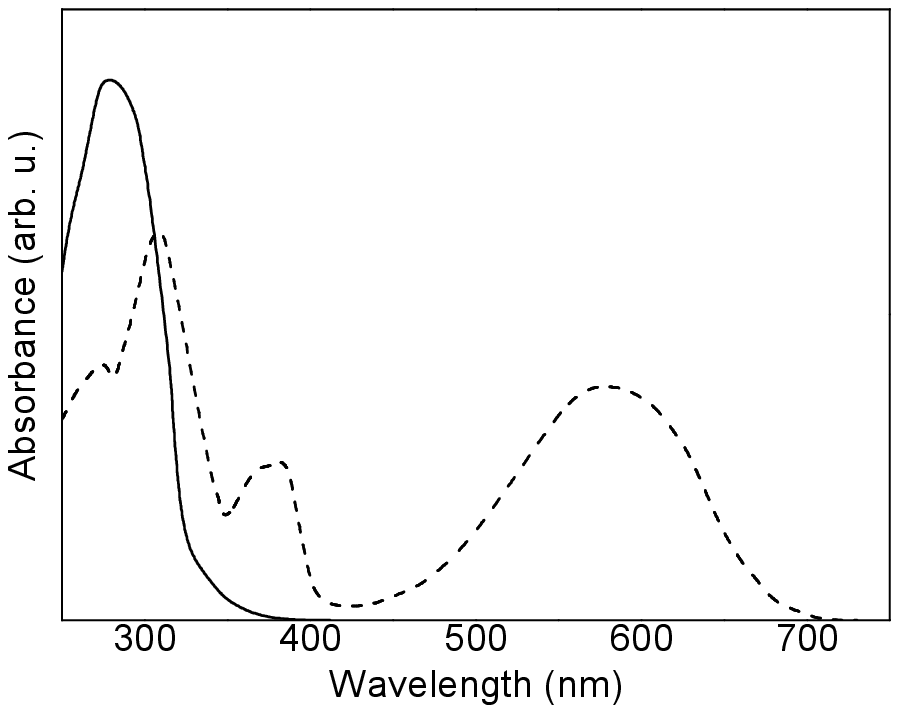}\\
{\large\it b}\vskip2mm
\includegraphics[width=8.7cm]{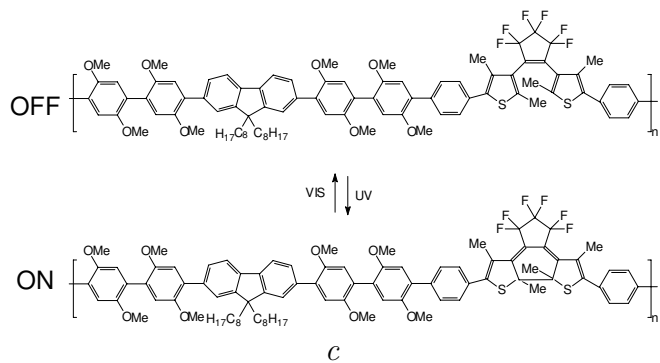}\\
{\large\it c} \vskip-3mm\caption{({\it a}) The photochromic reaction
in the diarylethene system. ({\it b}) Electronic absorption spectra
of the two forms of diarylethene: full line -- stable form; dashed
line -- metastable (colored) form. ({\it c}) Diarylethene
incorporated in the main chain of a $\pi $-conjugated polymer [11] }
\end{figure}

\begin{figure}
\includegraphics[width=8.7cm]{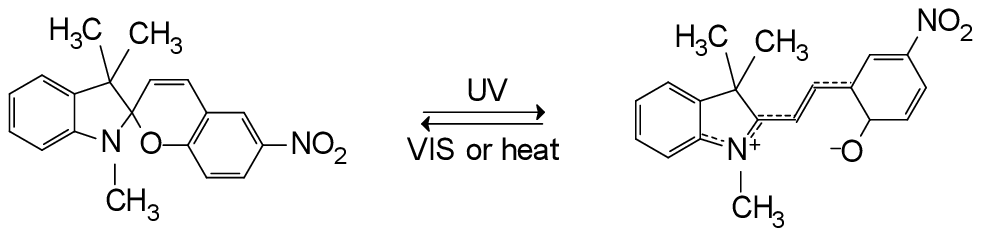}\\
{\large\it a}\\ \vskip3mm
\includegraphics[width=8cm]{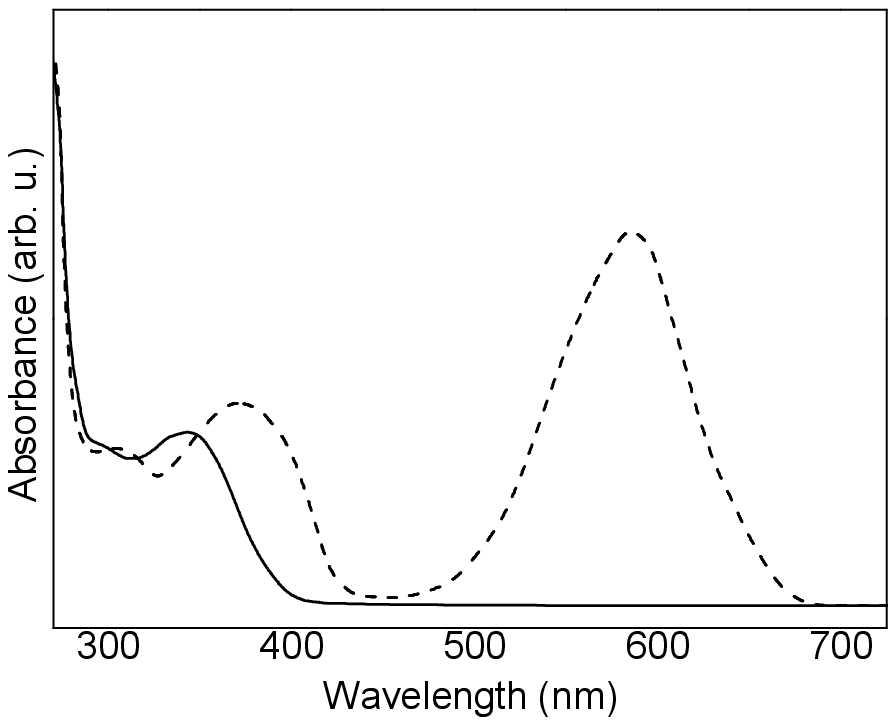}\\
{\large\it b}\vskip3mm
\includegraphics[width=8cm]{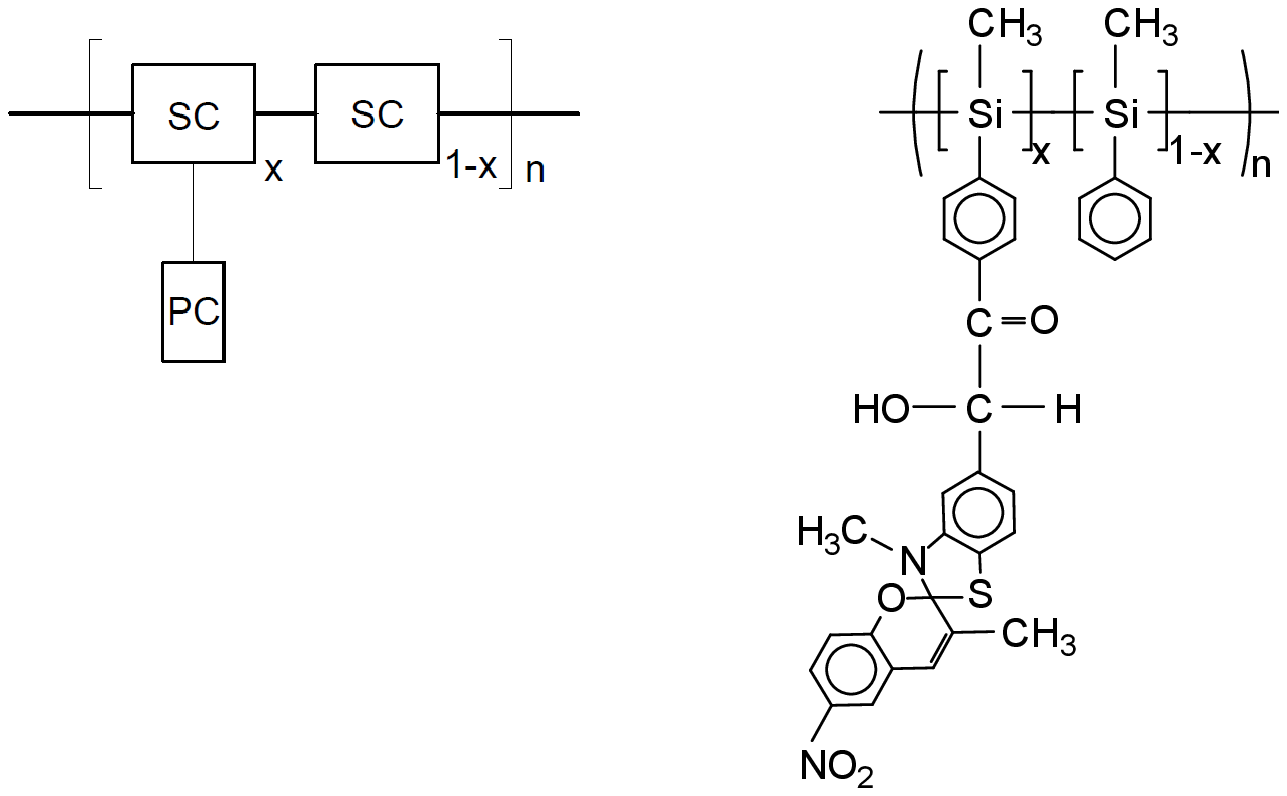}\\
{\large\it c} \vskip-3mm\caption{({\it a}) Photochromic reaction in
the spiropyran-merocyanine system. ({\it b}) Electronic absorption
spectra of the two forms of spiropyran: full line -- stable form
(spiropyran); dashed line -- metastable form (merocyanine). ({\it
c}) The architecture of a switch put forward in [3]: semiconducting
copolymer (segments SC) and photochromic side group PC, chemically
attached to the main chain (left panel), and its postulated
synthetic realization (right panel) }
\end{figure}

\begin{figure*}
\includegraphics[width=8.5cm]{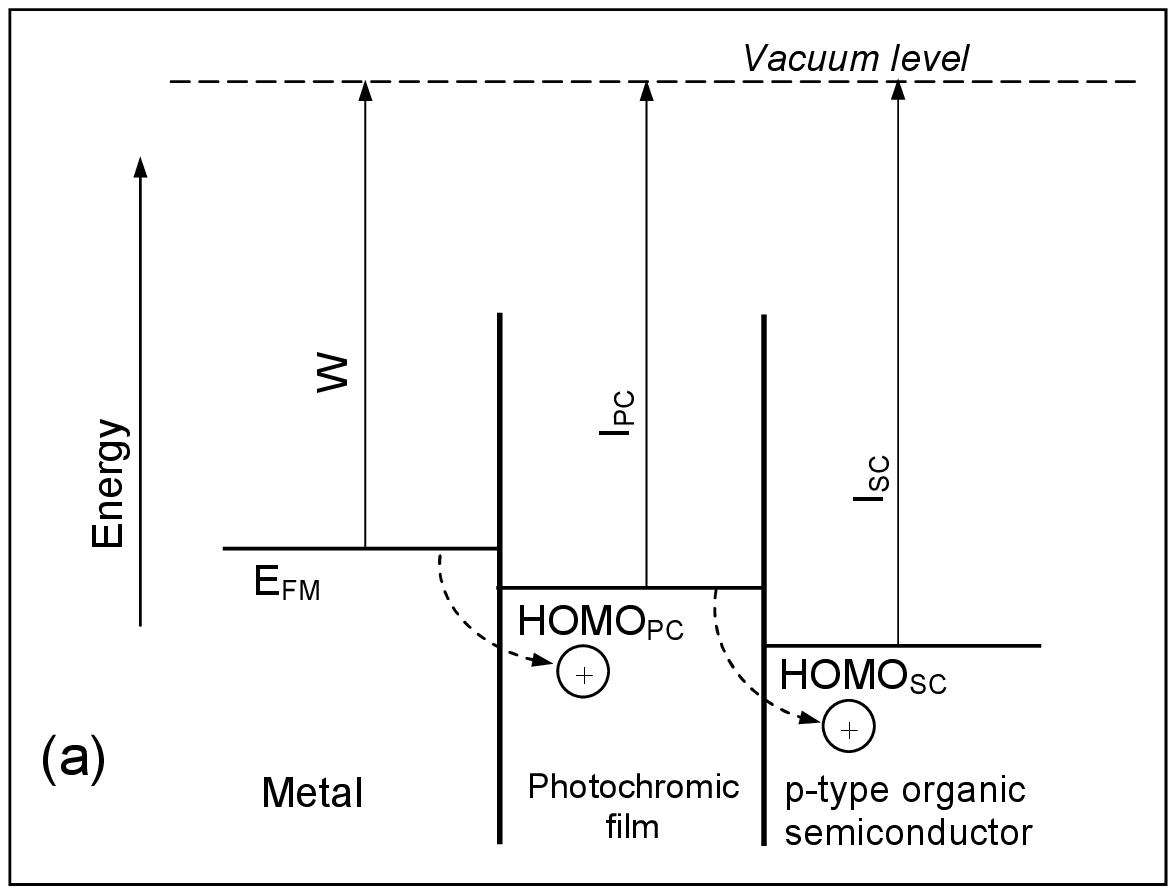}\hspace{0.5cm}\includegraphics[width=8.5cm]{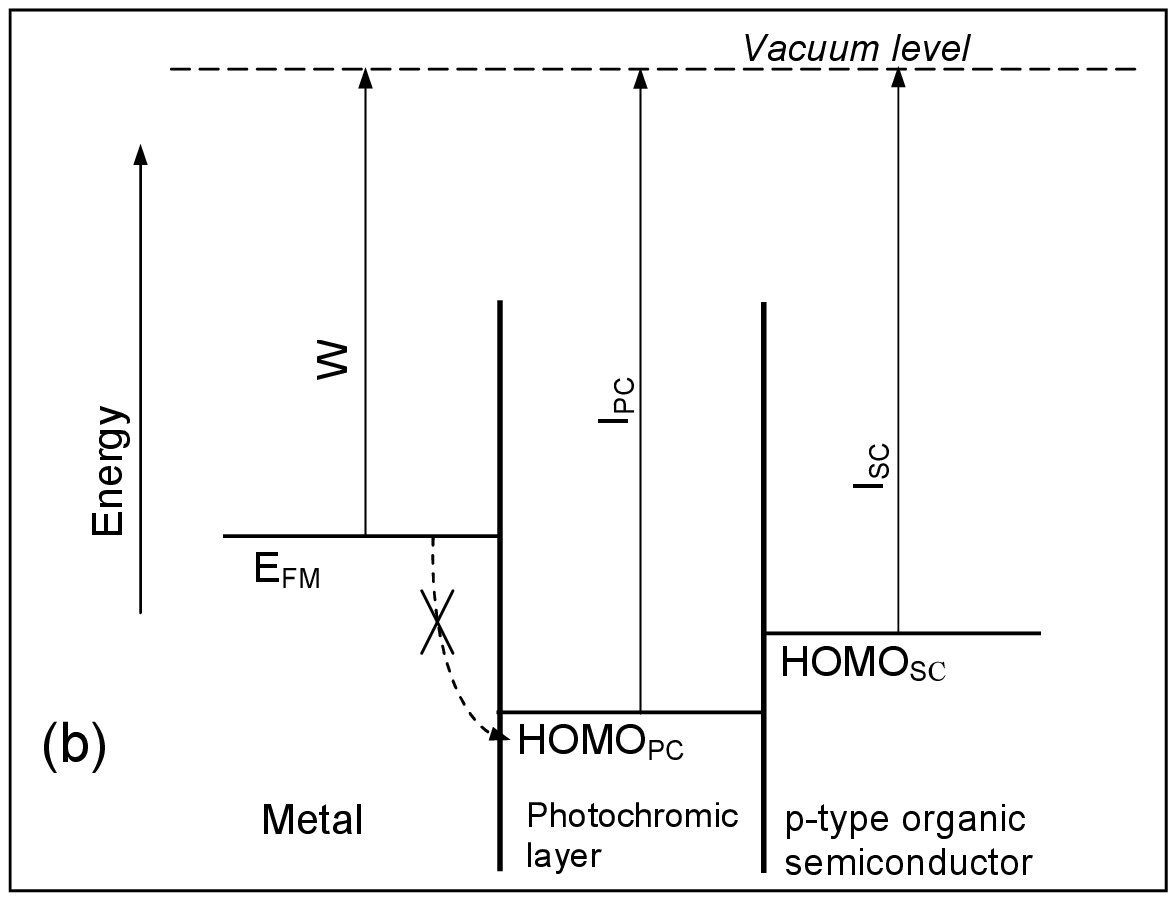}
\vskip-3mm\caption{ Energy diagrams for the hole injection from a
metal contact into a hole-transporting organic semiconductor through
a photochromic layer. ({\it a}) Due to a small energy difference
between the ionization energy $I_{\rm PC}$ of the stable form of the
photochromic system and the work function of a contact metal $W_{\rm
met}$, the hole injection and, consequently, the charge carrier
transport are possible. ({\it b}) The height of the interface
barrier between the metastable form and the metal is higher,
preventing the hole injection. $E_{\rm F,met}$ is the Fermi level of
the metal, $I_{\rm SC}$ is the ionization energy of the
hole-transporting organic semiconductor (adapted from [18]) }
\end{figure*}

{\it iv)} fast photochemical reaction with small conformational
changes of photochromic molecules -- large changes of a molecular
conformation make the photochromic reaction slow;

{\it v)} the matrix must have free volume enabling conformational
changes of the photochromic moieties;

{\it vi)} in the case of a binary system consisting of a
photochromic additive dissolved in a polymer or low-molecular weight
matrix, the mutual miscibility must be sufficiently high;

{\it vii)} in order to prevent the uncontrolled thermal reverse
reaction, the ground-state energy barrier between two forms of
the photochromic system must be sufficiently high;

{\it viii)} the difference of dipole moments and/or ionization
energies of both forms should be as large as possible.\looseness=1

It must be realized that the simultaneous fulfilment of all these conditions can
be quite difficult. Thus, the practical realization of a technologically useful
single-phase two-electrode optoelectrical switch is a task that requires
much effort, in particular from materials scientists and synthetic chemists.

\subsection{Photochromic film on the surface of an organic
semiconductor}

A different approach was proposed by Jakobsson {\it et al}. [18]. In
their structure (Fig. 5), the photochromic material was placed
between an organic semiconductor and a metal; the photochromic
reaction resulted in the reversible change of a barrier for
the injection of charge carriers, by reversibly modifying the current.
Considering a $p$-type semiconductor, the basic parameter is the
relation between energies of the HOMO levels (ionization energies if
measured with respect to the vacuum level) of stable and metastable
forms of the photochromic system and the work function of the metal
electrode. In some materials, the ionization energies of two states
of the photochromic moieties are quite different (as, for example,
is the case of diarylethene and spiropyran). A change of
the ionization energy modifies the energy difference between HOMO of the
photochromic material and the Fermi level of an electrode, resulting in
a change of the injection barrier for holes and, hence, the current
flow through the structure~[18].\looseness=1

\section{Photoswitchable Organic Field Effect Transistors}

Organic field effect transistor (OFET) is a three-electrode
structure (cf. Fig. 6), in which the current flowing in a
semiconductor between two active electrodes (``source'' and
``drain'') can be modified by a voltage applied to the third
electrode (''gate'') separated from the semiconductor by an
insulating film [19]. In general, the source-drain current, $I_{\rm
SD}$, is therefore a function of the source-drain voltage, $U_{\rm
SD}$, and gate voltage, $U_{\rm G}$. The expressions describing the
current-voltage characteristics read
\begin{equation}
I_{\rm SD}=\mu_{\rm FET}\frac{WC_{\rm ins}}{L}\left( U_{\rm
G}-U_{\rm T}-\frac{U_{\rm SD}}{2}\right)) U_{\rm SD},
\end{equation}
\begin{equation}
I_{\rm SD}=\mu_{\rm FET}\frac{WC_{\rm ins}}{2L}\left( U_{\rm
G}-U_{\rm T}\right)^2.
\end{equation}
Equation (4) describes the course of the characteristics for low
$U_{\rm SD}$ (``linear range''), whereas Eq. (5) is valid at high
$U_{\rm SD}$ (``saturation range''). In the above equations,
$\mu_{\rm FET }$ is the mobility of charge carriers, $U_{\rm T }$ is
the threshold voltage, $C_{\rm ins}$ is the insulator capacitance,
and $L $ and $W$ stand for the channel length and the width,
respectively.

In these structures, the photoswitchable molecules can be placed
either in the semiconductor or in the insulator. Experiments with
structures in the former architecture (Fig. 6,{\it a}) yielded
results resembling those obtained with similar systems in the
two-electrode arrangement [20]. The result can be easily
rationalized: the switching is expected to affect the mobility of
charge carriers in the same way as in the two-electrode
structures. Thus, the processes responsible for the switching, its
reversibility, its rate {\it etc}. are very much the same in the
materials under study irrespective of the experimental geometry.

\subsection{Photochromic species in the OFET insulator}

An alternative to the architecture discussed above would be the
use of a photoswitchable insulator (Fig. 6,{\it b}). This can be
realized by either using a photoactive insulator (e.g., an
insulating polymer with photochromic pendant groups) or by
dissolving photochromic molecules in a neutral insulator. The
mechanism(s) responsible for the process will be discussed below.

The structures depicted in Fig. 6,{\it b} were studied by Shen
{\it et al.} [21] and, independently, by Lutsyk {\it et al}. [22,
23]. The former group reported on a reversible switching in an
OFET containing $p$-type organic semiconductor pentacene, and
poly(methyl methacrylate) (PMMA) containing dissolved photochromic
spiropyran (Fig. 4,{\it a}), acting as photoswitchable gate
insulator. The switching was also reported in an OFET containing
$n$-type organic semiconductor perfluorinated Cu phthalocyanine
[21]. The authors attributed the effect to a reversible change of
the electric permittivity.

\begin{figure}
\includegraphics[width=7cm]{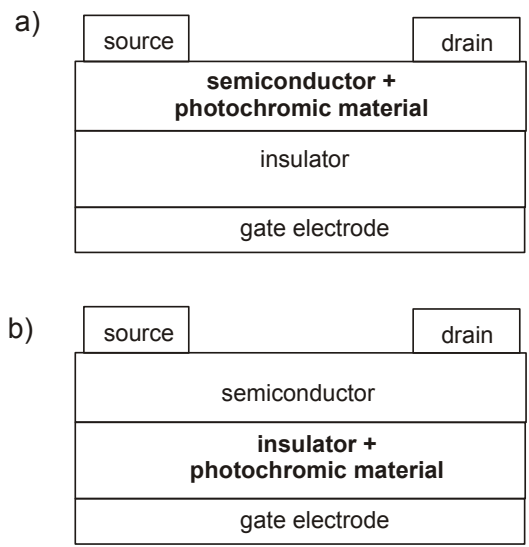}
\vskip-3mm\caption{ Two modifications of an OFET containing a
photochromic material: ({\it a}) the photochromic system admixed (or
chemically attached) to the semiconductor; ({\it b}) the
photochromic system admixed (or chemically attached) to the
insulator }
\end{figure}

\begin{figure}
\includegraphics[width=5.5cm]{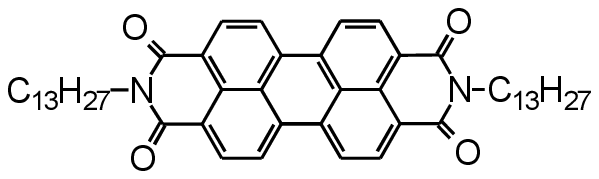}\\
{\large\it a}\\ \vskip2mm
\includegraphics[width=7cm]{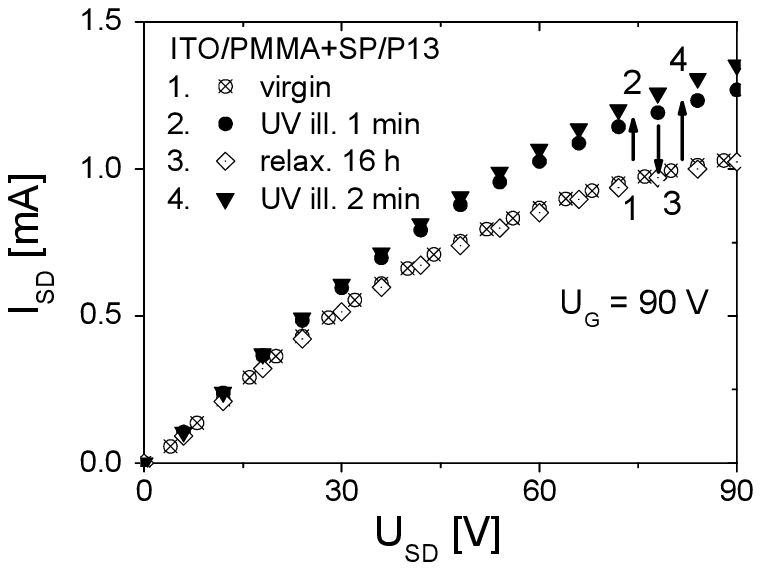}\\
{\large\it b}\\ \vskip2mm
\includegraphics[width=7cm]{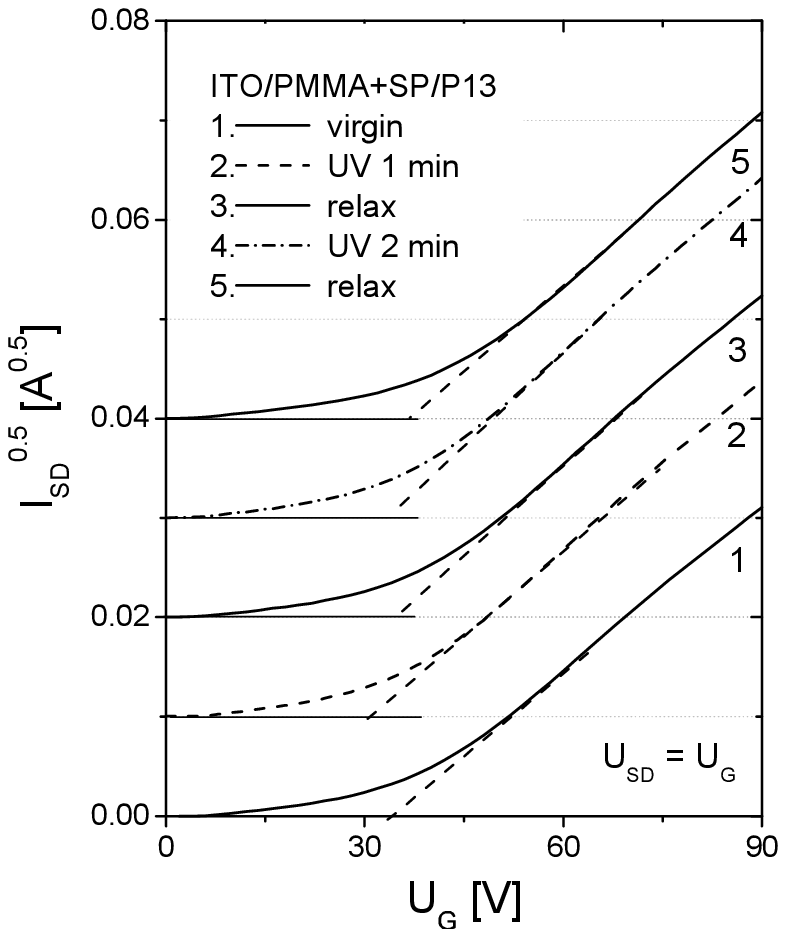}\\
{\large\it c} \vskip-2mm\caption{ Switching in the OFET consisting
of $n$-type organic semiconductor N,N$\prime
$-ditridecylperylene-3,4,9,10-tetracarboxylic diimide (P13), and
insulating PMMA containing 10 wt.{\%} of photochromic spiropyran
[22, 23]. ({\it a}) Chemical formula of the semiconductor. ({\it b})
An exemplary output current-voltage characteristics measured in the
system demonstrating the switching. The sequence of measurements is
indicated by arrows. ({\it c}) Locus ($U_{\rm SD} = U_{\rm G})$
current-voltage characteristics. The sequence of measurements is
indicated in the diagram. To facilitate the comparison, curves 2--5
have been vertically displaced. The broken lines show linear fits to
the experimental curves above the threshold voltages }
\end{figure}

Lutsyk {\it et al}. [22, 23] studied the photoswitching in the
structure consisting of an $n$-type organic semiconductor
tridecyl-substituted bis-perylenediimide (hereafter referred to as
P13 -- see Fig. 7,{\it a}) and PMMA containing dissolved spiropyran.
The results, shown in Fig.~7, demonstrate the effective switching
triggered by light. The photoswitching ratio was found dependent on
the gate voltage amounting to ca. 7--8 at the lowest voltages and
decreasing to ca. 1.3 at the highest voltages.

Two processes can account for the light-induced reversible modulation of
the current in the structures under consideration:

{\it i)} reversible change in the capacitance of the insulator;

{\it ii)} reversible modification of the insulator-semiconductor
interface.

Both processes stem from a reversible change of the dipole moment of
photochromic species (admixed molecules or side groups chemically attached
to the polymer chain) located either in the bulk of the insulator or at the
semiconductor-insulator interface.

\subsubsection*{3.1.1. Photochromic dipoles at the \\ \phantom{3.3.1. }insulator-semiconductor interface}

The importance of switchable polar species located on the surface of
an organic semiconductor has been discussed in several publications.
Shen {\it et al}. [24] deposited photochromic spiropyran molecules
on the surface of $p$-type and $n$-type semiconductors in OFET. In
$p$-type transistors, an increase of the source-drain current after
the UV illumination and a decrease of the current after visible
light SP was observed, whereas the opposite effect was detected in
$n$-type devices: the current decreased after the UV illumination
and rose after the exposure to visible light. The effect has been
explained by assuming the contribution of a local (negative) gate
voltage yielded by the dipoles created upon irradiation, resulting
in an increase of the current in $p$-type devices, and to a decrease
of the current in $n$-type OFETs. Spiropyran grafted to
single-walled carbon nanotubes in transistors also showed the
photoswitching effect [25]: the current dropped after the UV
illumination. The phenomenon has been explained by the creation of
dipole traps for charge carriers by highly polar photomerocyanine
molecules produced by the photochromic reaction. However, spiropyran
grafted to polyaniline chains showed the opposite photoswitching
behavior: an increase of the current under the UV irradiation. The
effect has been explained assuming the contribution from a local
field, complementary to the gate field, appearing due to the
creation of a highly polar form of the photochromic
system~[26].

Recent reports on the reversible switching of the current in an OFET
containing PMMA insulator doped with spiropyran also pointed to the
effect of photochromic molecules located at the interface [22, 23].
Due to a strong interaction of interface dipoles with charge
carriers in the channel (Eq. 2), interfacial charged states can be
formed. A fraction of photochromic molecules dispersed in the
insulator situated at the insulator--semiconductor interface
creates local states on adjacent semiconductor molecules [15, 24].
The semiconductor molecules situated in a vicinity of suitably
oriented dipoles (i.e., those for which $E_{\rm id}  < 0$) will act
as dipolar traps [15]. The remaining ones may only scatter
the approaching carriers. In the first approximation, the depth of a
dipolar trap will be equal to the charge-dipole interaction energy,
i.e., will be described by Eq. (2). Let us consider an example:
photochromic spiropyran (Fig. 4,{\it a}) admixed to PMMA. In its
stable form, spiropyran (SP) has the dipole moment of ca 5.4 D.
The illumination with UV light converts the stable form into metastable
photomerocyanine (MR) having the dipole moment of ca. 11 D [6].
Taking the distance from the interfacial dipoles to the first
molecular layer of the semiconductor equal to ca. 0.35 nm, the maximum
depth of dipolar traps, given by Eq. (2) with $\cos \Theta = - 1$,
estimated for SP dipoles is not higher than 0.35 eV, increasing to
ca. 0.7 eV after the conversion of SP to merocyanine by the UV
illumination.

Changes in the parameters characterizing the interface traps affect
the threshold voltage ($U_{\rm T})$. The threshold voltage is
related to the parameters of the semiconductor and the insulator by the
equation [19]
\begin{equation}
U_{\rm T}=\Phi_{\rm s}+\Phi_{\rm FG}+\frac{Q_{\rm s}}{C_{\rm ins}},
\end{equation}
where $\Phi _{\rm s}$ stands for the surface potential of the
semiconductor, $\Phi _{\rm FG}$ is a potential supplied by the
floating gate, and $Q_{\rm s}$ is a sum of the space charge in the
semiconductor and a charge accumulated in the interface states.
Consequently, one may expect changes in $U_{\rm T}$ due to
surface potentials and charges and the bulk capacitance. It comes
from the above equation that an increase in the depth and/or density
of traps should result in an increase in $Q_{\rm s}$ and/or $\Phi
_{\rm s}$, and in an increase in $U_{\rm T}$. The bulk effect will
be tackled in the subsequent section.

The presence of the interface traps affects also the shapes of the
current-voltage characteristics in the subthreshold region (i.e.,
for $U_{\rm G}< U_{\rm T})$ [19]. The current-voltage dependences in
the subthreshold range of voltages are usually characterized by the
subthreshold swing describing the slope of the characteristics:
\begin{equation}
S=\left [\frac{\partial(\log_{10}I_{\rm SD})}{\partial U_{\rm
G}}\right ]^{-1}_{U_{\rm G}<U_{\rm T}}.
\end{equation}
According to [19], the latter parameter is given by the equation
\begin{equation}
S=\frac{kT}{e}\ln\left [ 10\left (1+\frac{C_{\rm D}+C_{\rm
IT}}{C_{\rm ins}}\right )\right ],
\end{equation}
where $C_{\rm D}$ is the capacitance of the depletion layer of the
semiconductor, $C_{\rm ins}$ is the capacitance of the insulator,
and $C_{\rm IT}$ is the capacitance associated with the interface
traps. It comes from Eq. (8) that, at 300 K, the lowest available
value of $S$ amounts to ca. 60 mV/decade; the experimentally
obtained values of the parameter are usually higher by a half to one
order of magnitude. Two parameters appearing in Eq. (8), $C_{\rm
ins}$ and $C_{\rm IT},$ can be reversibly modified by the
photochromic reaction, thus influencing the slope of the current--voltage
characteristics. The conclusion finds a support in the
results of calculations published by Scheinert {\it et al.} [27] who
performed simulations of the I-V characteristics in an OFET showing
that the introduction of traps at the semiconductor-insulator interface
results in an increase of the current in the subthreshold region and
practically does not affect the current at $U_{\rm G} >  U_{\rm T}$.
The increase of the subthreshold current was found dependent on the
density and the depth of interface states [27]. Changes in the
subthreshold currents during the reversible conversion of spiropyran
into merocyanine, reported in an earlier paper of the present
authors [23], have been explained by the transformation of shallow
dipolar traps associated with spiropyran molecules into deep
interfacial states associated with merocyanine molecules. One can
thus infer that, in OFETs with polar dopants admixed to insulators,
a contribution of interface dipoles to the modulation of
the source-drain current may be noticed at gate voltages below the
threshold voltage.

\subsubsection*{3.1.2. Photochromic dipoles in the insulator bulk }

It follows from Eqs. (4) and (5) that, with other parameters kept constant,
the source-drain current is proportional to the capacitance of the
insulator. Thus, a reversible change in \textit{$\varepsilon $} should result in an appropriate
change in the current.

A semiquantitative model describing the dependence of \textit{$\varepsilon $} on the dipole moment of
polar molecules dispersed in a non-polar dielectric matrix was put forward
in [23]. The equation developed from the Onsager theory of dielectrics [28]
was written in the form
\begin{equation}
\frac{(\varepsilon -1)(2\varepsilon
+1)}{\varepsilon}=A+x(B+Dm^2_{\rm phot}),
\end{equation}
where $x$ is the mole fraction of polar photochromic molecules,
$m_{\rm phot }$ is their dipole moment, and $A$, $B,$ and $C$ stand
for parameters independent of $x$ and $m_{\rm phot}$. As expected,
the electric permittivity of an insulator containing polar molecules
should increase with increase in the dipole moment of a dopant and its
concentration. However, the predicted changes in $\varepsilon $
significantly (by ca. one order of magnitude) exceed the
experimental ones. Apart from the deficiencies of the simple model
developed in [23], the explanation of the discrepancy should be
sought in material factors such as the aggregation of polar species
markedly decreasing their effective dipole moment, incomplete
conversion of the stable form of the photochromic system, {\it etc.}
It should also be noted that a similar discrepancy was reported by
Shen {\it et al.} [21] who estimated changes in \textit{$\varepsilon
$} from molecular dynamics calculations.

Changes in the electric permittivity should also affect the
threshold voltage via changes in the insulator capacitance: it comes
from Eq. (6) that an increase in $\varepsilon $ (i.e., an increase
in the capacitance) should result in a {\it decrease} in the
threshold voltage, contrary to the trend expected of the surface
effect. The $U_{\rm T}(\varepsilon )$ dependence reported in [23]
seems to indicate that the bulk effect plays a decisive role in
OFETs comprising PMMA/spiropyran as an insulator and perylene
derivative as a semiconductor. $U_{\rm T}$ shifted to lower values
after the transformation of spiropyran to a highly polar form resulting to
an increase of \textit{$\varepsilon $} of the insulator.

\section{Final Remarks}

The models described in the present paper and the results of
numerous experiments point to a feasibility of construction of a
optoelectric switch whose heart would be a photoswitchable
molecular system. Various architectures of such devices have been
presented; it seems that the three-electrode systems are more
suitable, as they allow one to decouple the light-controlled
chemical reaction from the transport of charge carriers.
Nevertheless, quite stringent conditions must be met regarding the
rates of the photochromic conversion and the chemical stability of
a system before the devices can be considered usable in
technological applications. Thus, the matter requires a concerted
effort of synthetic chemists, materials scientists, and engineers.

\vskip3mm This work was supported by the European Commission through
the Human Potential Programme (Marie-Curie RTN BIMORE, Grant
No.\,MRTN-CT-2006-035859), and by the Wroc{\l}aw University of
Technology.

\rezume{%
БІСТАБІЛЬНІ~~ ОРГАНІЧНІ~~ МАТЕРІАЛИ\\ В ОПТОЕЛЕКТРИЧНИХ ПЕРЕМИКАЧАХ:\\
ДВОЕЛЕКТРОДНІ ПРИСТРОЇ\\ ТА ОРГАНІЧНІ ПОЛЬОВІ\\ ТРАНЗИСТОРИ }{Ю.
Свораковскі, П. Луцик} {У даній роботі наведено короткий огляд
дослідження властивостей органічних\rule{0pt}{9pt} матеріалів та
структур, що можуть бути використані\rule{0pt}{9pt} в
оптоелектричних перемикачах, тобто перемикачах, в яких
зміна\rule{0pt}{9pt} електричних властивостей відбувається за
рахунок освітлення\rule{0pt}{9pt} відповідної довжини хвилі. При
цьому описано структури,\rule{0pt}{9pt} що діють завдяки реверсивній
фотохімічній\linebreak реакції, яка\rule{0pt}{9pt} відбувається у
фотохромних молекулярних матеріалах.\rule{0pt}{9pt}}

\end{document}